\documentclass[useAMS,usedcolumn]{mn2e}
\usepackage{psfig}

\newif\ifAMStwofonts

\newcommand{\bq}[1]{\begin{equation} \label{#1}}
\newcommand{\eq}{\end{equation}}

\title{$RXTE$ Observations of 3C~273 between 1996 and 2000:\\
 Variability Timescale and Jet Power}

\author[Kataoka et al.]
       {Jun Kataoka$^{1,2}$, Chiharu Tanihata$^{3,4}$, Nobuyuki Kawai$^{1,2}$,
Fumio Takahara$^5$,
\newauthor
Tadayuki Takahashi$^{3,4}$, Philip G. Edwards$^3$, and Fumiyoshi Makino$^2$\\
\vspace{2mm}\\
$^1$ Department of Physics, Faculty of Science, Tokyo Institute of
	Technology, Meguro-ku, Tokyo, Japan\\
$^2$ National Space Agency of Japan, Tsukuba, Ibaragi, Japan\\
$^3$ Institute of Space and Astronautical Science, Sagamihara,
Kanagawa, Japan\\
$^4$ Department of Physics, Faculty of Science, University of Tokyo, 
Tokyo, Japan\\
$^5$ Department of Earth and Space Science, Toyonaka, Osaka, Japan}

\date{Accepted 2002 *****
      Received 2002 *****;      in original form 1988 October 11}

\pagerange{\pageref{firstpage}--\pageref{lastpage}}
\pubyear{1994}

\begin{document}

\maketitle

\label{firstpage}

\begin{abstract}

We present the results of a long-look monitoring of 3C~273 with $RXTE$
between 1996 and 2000. A total of 230 observations amounts to a net
exposure of 845~ksec, with this spectral and  variability 
analysis of 3C~273 covering the longest
observation period available at hard X-ray energies. Flux variations
by a factor of 4 have been detected over 4~years, whereas less than 30$\%$
flux variations have been observed for individual flares on time-scales
of $\sim$\,3~days. Two temporal methods, the power spectrum density (PSD)
and the structure function (SF), have been used to study the variability
characteristics of 3C~273.
The hard X-ray photon spectra generally show a power-law shape
with a differential photon index of $\Gamma$$\simeq$ 1.6$\pm$0.1.
In 10 of 261 data segments, exceptions to power-law behaviour have been 
found: (i) an additional soft excess below 4~keV, and (ii) a broad Fe 
fluorescent line feature with $EW$ $\sim$ 100$-$200 eV.
Our new observations of these previously reported X-ray features may 
imply that 3C~273 is a unique object whose hard X-ray emission 
occasionally contains a component which is not related to a beamed 
emission (Seyfert like), but most hard X-rays are likely to originate 
in inverse Compton radiation from the relativistic jet (blazar like).
Multi-frequency spectra from radio to $\gamma$-ray are presented in 
addition to our $RXTE$ results.
The X-ray time variability and spectral evolution are
discussed in the framework of beamed, synchrotron self-Compton
picture. We consider the ``power balance'' (both radiative and
kinetic) between the accretion disk, sub-pc-scale jet, and the 10~kpc-scale
jet.

\end{abstract}


\section{Introduction}

As the brightest and nearest ($z$ = 0.158) quasar, 3C~273 is the
ideal laboratory for studying active galactic nuclei (AGN). 
Studies of this source are relevant
to all AGN physics, as 3C~273 displays significant flux variations,
has a well-measured wide-band spectral energy distribution, and
has a relativistic jet originating in its central core 
 (see Courvoisier 1998 for a review).
VLBI radio observations of the pc-scale jet have
shown a number of jet components moving away from the core at velocities
apparently faster than the speed of the light
(e.g., Pearson et al.\ 1981; Vermeulen \& Cohen 1994).
The collimated jet structure extends up to $\sim$\,50~kpc from the core.
Since 3C~273 is bright at all wavelengths and on various scale sizes,
it provides a valuable opportunity to probe the most inner
part of the accretion disk ($\sim$\,10$^{-4}$~pc) as well as the large
scale jet ($\sim$\,10$^4$~pc) at the same time.

3C~273 is generally classified as a blazar 
and is also a prominent $\gamma$-ray source.
It was the only extra-galactic source of 
gamma-rays identified in COS-B observations (Swanenburg et al. 1978), 
and was subsequently
detected at energies in the 0.05\,MeV to 10\,MeV range with OSSE
(McNaron-Brown et al. 1995), in the 0.75\,MeV to 30\,MeV range with
COMPTEL (Sch\"onfelder et al. 2000), and above 100\,MeV in numerous EGRET
observations (Hartman et al. 1999).
EGRET observations helped establish that the overall spectra of
blazars (plotted as $\nu$$F_{\nu}$) have two pronounced continuum
components: one peaking between IR and X-rays, and the other in the
$\gamma$-ray regime (e.g., Mukherjee et al. 1997). The low energy component is
believed to be produced by synchrotron radiation from relativistic
electrons in magnetic fields, while inverse-Compton scattering by the
same electrons is thought to be dominant process responsible for the
high energy $\gamma$-ray
emission (Ulrich, Maraschi, \& Urry 1997). The radiation is
emitted from a relativistic jet, directed close to our line of sight
(e.g., Urry \& Padovani 1995).

3C~273 is no exception to this picture, but an additional $Big$ $Blue$ $Bump$ 
(hereafter $BBB$\,) dominates the optical--soft-X-ray emission 
(see Paltani, Courvoisier, \& Walter 1998 ; Robson 1996 for a
review). Interestingly, although similar excesses
have been observed in the optical-UV region of Seyfert galaxies,
they have not been reported for any blazar other than 3C~273. Although its
origin is still far from being understood, it has been proposed that
$BBB$ may be due to thermal emission from the surface of a
standard accretion disk (Shields 1978), including optically thin parts of
the disk and a corona (see Czerny 1994 for a review). Various
models have been suggested (e.g., Courvoisier \& Clavel 1991), but it is
generally agreed that the $BBB$ is nearly isotropic emission
from the vicinity of the central black hole, presumably from the
accretion disk. 

The presence of a fluorescent emission line at 6.4\,keV
is a signature of X-ray reprocessing by cold material.
There is some evidence for the presence of a weak line at this energy in
the X-ray spectrum of 3C~273. One of the $Ginga$  observations showed 
evidence for the line at 99$\%$  level, but the other observations
provided only upper limits (Turner et al. 1990). A line detection is
also reported by Grandi et al. (1997) and Haardt et al. (1998) in a 
$BeppoSAX$ observation in 1996. $ASCA$ provided only upper limits in
a 1993 observation (Yaqoob et al. 1994), whereas a broad line feature was 
clearly detected in 1996 observations. This somewhat confused
situation probably means that both thermal (as for Seyferts) 
and non-thermal (as for blazars) emission processes are taking place in 
this particular object. However, it is completely unknown (i) which 
process dominates the radiation and (ii) how often the Fe line is
``visible'' in the photon spectrum.

Many $\gamma$-ray blazars, including 3C~273, have shown 
large flux variations either, on time-scales as short as a day for some
objects
(e.g., von Montigny et al. 1997; Mukherjee et al. 1997). 
However, there are only a few blazars whose variability characteristics are 
well studied in the high energy bands (both X-ray and $\gamma$-ray bands).  
For example, recent X-ray studies of Mrk 421, the proto-typical 
``TeV emitting'' blazar,  have revealed that 
(i) the variability time scale is $\sim$ 1~day, and 
(ii) the flux variation in the X-ray and the TeV $\gamma$-ray bands is well 
     correlated on time-scales of a day to years
(see Takahashi et al. 2000; Kataoka  et al. 2001). Clearly, these 
observations provide important clues to understanding jet physics, and 
potentially for discriminating between various emission models for blazars. 
Although 3C~273 is a bright object and particularly well-sampled, 
it has not yet been possible to undertake such studies since data from 
previous X-ray satellites were too sparse (e.g., 
the 13 observations spanning over 5 years of Turner et al. 1990).

In this paper, we analyze the archival hard X-ray data obtained with
$RXTE$ between 1996 and 2000, with a total exposure of 845~ksec.
This report of both the temporal
and spectral variability
of 3C~273 is thus based on the highest quality and most densely sampled
data in this energy band.
The observation and data reduction are described in
$\S$2. Two temporal methods are introduced;
the power spectrum density ($\S$3.1)
and the structure function ($\S$3.2).
The X-ray spectral evolutions are summarized in $\S$$\S$4.1
and 4.2. Multi-frequency spectra are presented in $\S$4.3.
In $\S$5, we discuss scenarios which systematically account for
the hard X-ray variability and spectral evolution of 3C~273.
Throughout the paper, we discuss the energetics between the
central engine and the relativistic
jet, taking into account observations of the large scale jet by
$Einstein$, $ROSAT$ and $Chandra$.
Finally, in $\S$6, we present our conclusions.

Throughout this paper, we adopt $H_0 = 75$~km\,s$^{-1}$\,Mpc$^{-1}$
and $q_0 = 0.5$. The luminosity distance to the source is
$d_{\rm L} = 2.02 \times 10^{27}$~cm.

\section{Observation and Data Reduction}

3C~273 was observed 230 times with the X-ray satellite $RXTE$ between
1996 February and 2000 February, with a net exposure of 845~ksec.
The observations are summarized in Table~1. All $RXTE$ observations were
performed with Good Xe-16s plus Standard 1/2 modes for the Proportional
Counter Array (PCA; Jahoda et al.\ 1996).
The source counts were extracted from three Proportional Counter Units
(PCU0/1/2) for the 1996$-$1999 observations, which had much larger and less
interrupted exposures than those for PCU3 and PCU4. After 1999 May,
PCU1 experienced the same ``discharge problems'' as for PCU 3 and 4,
and was often operated with a
reduced high voltage supply. We thus use the data from two PCUs  (PCU0/2)
for the analysis of 1999 May to 2000 Feb.\ data (119~ksec of P40176).

We used only signals from the top layer (X1L and X1R) in order to obtain
the best signal-to-noise ratio. Standard screening procedures were
performed on the data, using analysis software package HEASOFT~5.0
provided by NASA/GSFC. Backgrounds were estimated using $pcabackest$
(version~2.1b) for the PCA, and subtracted from the data.
We have not used data from the
High Energy X-ray Timing Experiment (HEXTE) on board $RXTE$,
for two reasons; (i) the typical exposure for $RXTE$ observations was too
short to yield meaningful hard X-ray data of this variable source above
20~keV, and (ii) calibration problems make the analysis results quite
uncertain. Similar problems
exist
for the All Sky Monitor (ASM) data on board $RXTE$.
Given the large systematic errors already known
(e.g., enhanced noise at certain solar angles) in ASM faint source data,
we cannot justify the use of ASM data for detailed temporal studies.

\begin{table*}
\centering
 \begin{minipage}{150mm}
  \caption{Observation log of 3C~273.}
  \begin{tabular}{@{}llrrlll@{}}
\hline
Obs.ID &  Obs.date  & Obs.times\footnote{Number of observations
   conducted for each proposals.}  & Exposure (ksec) & Data
   Mode\footnote{S1; Standard 1, S2; Standard 2, GX1; Good Xe-1 16s, GX2; Good Xe-2 16s.}\\
\hline
P10330\footnote{4 observations with large offset angle ($\ge$ 0.5 deg)
   were not used for   the analysis.} & 1996 Jul 16 23:15$-$1996 Jul 18 13:21 & 6    & 116.6 & S1/S2/GX1/GX2\\
P10354 & 1996 Feb 02 01:12$-$1996 Aug 24 01:30 & 28   & 23.7  & S1/S2/GX1/GX2 \\
P20349 & 1996 Nov 03 15:30$-$1997 Dec 23 22:57 & 132  & 133.7 & S1/S2/GX1/GX2 \\
P30805 & 1998 Jun 24 07:44$-$1998 Jun 26 09:05 & 3    & 33.2  & S1/S2/GX1/GX2 \\
P40176 & 1999 Jan 04 06:25$-$2000 Feb 24 09:24 & 57   & 473.3 & S1/S2/GX1/GX2 \\
P40177 & 1999 Jan 19 18:32$-$1999 Feb 01 20:15 & 4    & 64.6  & S1/S2/GX1/GX2 \\
\hline
total & & 230    & 845.1  &  \\
\hline
\end{tabular}
\end{minipage}
\end{table*}

\begin{figure*}
\psfig{file=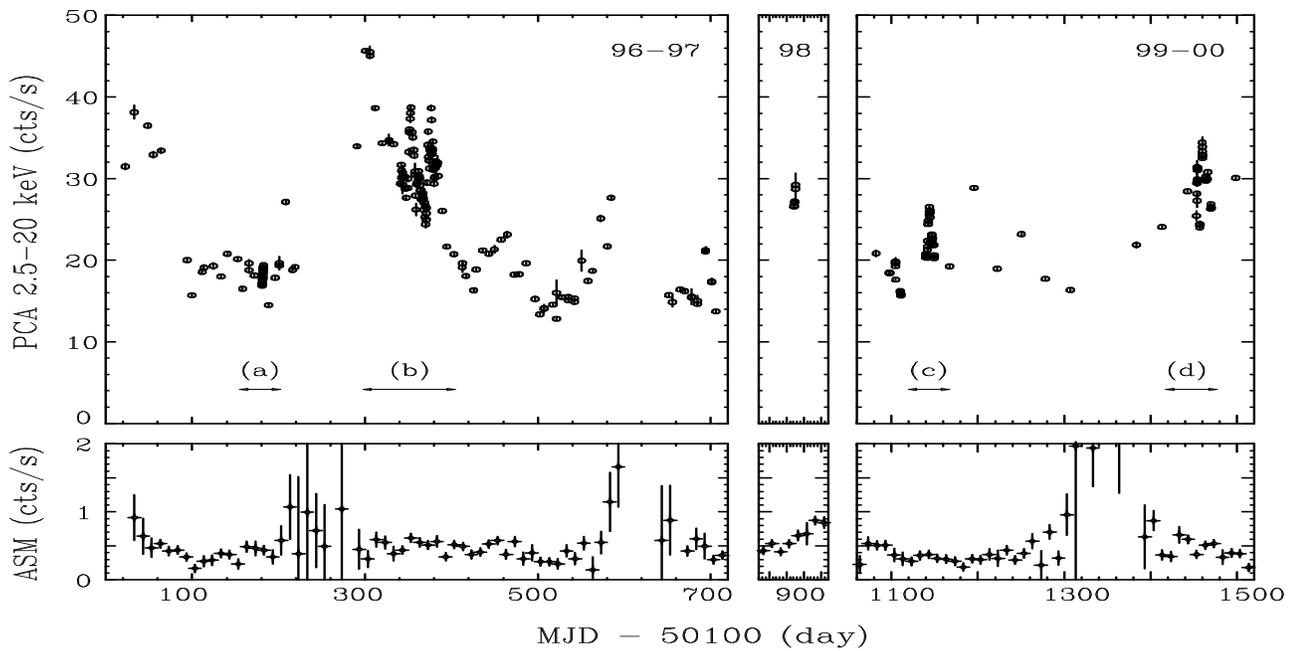,width=17cm,height=8.5cm}
\caption[h]{$Upper$; long-term light curve of 3C~273. The count rates of
 the top layer from $RXTE$ PCU 0/2 detectors are summed in the
 2.5$-$20~keV energy range.
 $Lower$; the count rates of All Sky Monitor (ASM) on-board $RXTE$.
 The data are binned in 10~day intervals.}
\label{fig1}
\end{figure*}

In order to obtain the maximum photon statistics and the best
signal-to-noise ratio, we selected the energy range 2.5$-$20~keV for the
PCA. The overall light curves are shown in Figure~1. The
ASM light curve (2--10 keV rates), in 10-day bins, is shown for 
comparison with the
PCA flux variations (though one finds no clear correlation between
them, because of the ASM problems described above).
We have used a bin size of 5760~sec (approximately the $RXTE$ orbital
period) for the PCA plot.  The 2PCU (PCU0/2) count rates are plotted
for all data from 1996$-$2000. Expanded plots of the segments (a)-(d),
indicated by arrows in Figure~1, are shown in Figure~2.
An intensive hard X-ray monitoring for more than a month was conducted
in 1996 simultaneously with IR telescopes (segment (b) in the Figures~1 and 2).
Results from this campaign have been
discussed by McHardy et al. (1999).

\section{Temporal Study}
\subsection{Power Spectrum Density}

A power spectrum density (PSD) analysis is most commonly used to
characterize the source variability. $RXTE$ data presented
in this paper are the highest quality data ever reported in the
X-ray band, enabling us to determine the PSD over a wider frequency
range than attempted previously. An important issue, however, is the
data gaps, which are unavoidable for low-orbit X-ray satellites.
Moreover, we must take great care since the observations were
sparsely scheduled; a net exposure of 845~ksec
corresponds to only 0.7$\%$ of the total span of 4~years.
In order to reduce the effect of data sampling, we follow a technique for
calculating the PSD of unevenly sampled light curves
(e.g., Hayashida et al.\ 1998).

\begin{figure}
\psfig{file=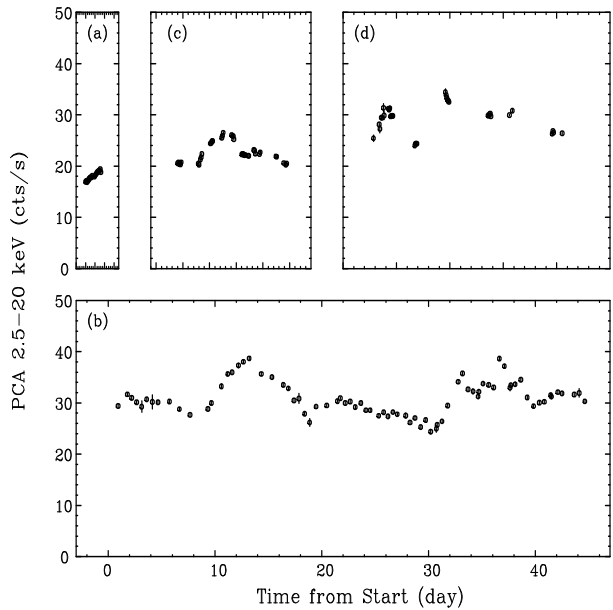,width=8.0cm,height=8.0cm}
\caption[h]{Expanded plots of Figure~1. The light curves are separately
 given for (a)-(d). The scale of time axis is same in all plots.}
\label{fig2}
\end{figure}

The $normalized$ PSD (NPSD) at frequency $f$ is defined as
\begin{eqnarray}
P(f) = \frac{[a^2(f)+b^2(f)-\sigma^2_{\rm stat}/n]T}{F_{\rm av}^2},\nonumber \\
a(f) = \frac{1}{n}\sum_{j=0}^{n-1} F_j {\rm cos} (2\pi f t_j), \nonumber \\
b(f) = \frac{1}{n}\sum_{j=0}^{n-1} F_j {\rm sin} (2\pi f t_j), \nonumber \\
\end{eqnarray}
where $F_{j}$ is the source count rate at time $t_j$
(0$\le$$j$$\le$$n-1$), $T$ is the data length of the time series, and
$F_{\rm av}$ is the mean value of the source counting rate. The power
due to photon-counting statistics is given by $\sigma_{\rm stat}^2$.

To calculate the NPSD of our data sets, we made light curves of three
different bin sizes for $RXTE$ data (256, 5760, and 43200~sec).
We then divided each light curve into ``segments'', which are defined as
continuous parts of the light curve. If the light curve contains a time-gap
larger than twice the bin size, we cut the light curve into two
segments, one each side of the gap. We then calculate the power at
frequencies $f$ = $k$/$T$ (1 $\le$ $k$ $\le$ $n$/2) for each segment
and take their average.  The lowest frequency end ($\sim$\,10$^{-6}$~Hz)
is about half the inverse of the longest continuous
segments, which is determined by the data in Figure~2(b) in our case.

Figure~3 shows the NPSD calculated using this procedure. These NPSD data
are binned in logarithmic intervals of $\sim$0.2 
(i.e., factors of approximately 1.6, as dictated by the data) to reduce
the noise. The error bars represent the standard deviation of the
average power in each rebinned frequency interval. The expected noise
power due to counting statistics, $\sigma_{\rm stat}^2$$T$/$n$$F_{\rm
av}^2$, is shown in Figure~3 as a dashed line.
As the NPSD has very steep power-law slope,
with the NPSD decreasing as frequency increases,
$no$ detectable variability exists in the
light curve for $f$ $\ge$ 10$^{-4}$~Hz (i.e., consistent with zero
power for short time variability $\le$ 10$^4$~sec).

A single power-law function is $not$ a good representation
of the NPSD, as the power-law fit above $10^{-5}$~Hz is too steep
for the data below $10^{-5}$~Hz ($\chi^2$=14.7 for 8~dof;
$P(\chi^2)=6\%$). A better fit was obtained using a broken
power-law function, where the spectrum is harder below the break.
The fitting function used was  $P(f)$ $\propto$ $f^{- \alpha_{\rm L}}$
for $f$ $\le$ $f_{\rm br}$ and
$P(f)$ $\propto$ $f^{- \alpha}$ for $f$ $\ge$ $f_{\rm br}$. The goodness
of the fit was significantly improved; $\chi^2=7.3$ for 6~dof,
$P(\chi^2)=30\%$. The best fit parameters are
$\alpha_{\rm L}=1.4\pm0.2$, $\alpha=2.6\pm0.1$, and
$f_{\rm br}=(4.7\pm1.5)\times10^{-6}$~Hz, respectively.
This corresponds to a variability time-scale for individual flares
of $t_{\rm var} \sim 3$~days, as can be seen in Figure~2.

\begin{figure}
\psfig{file=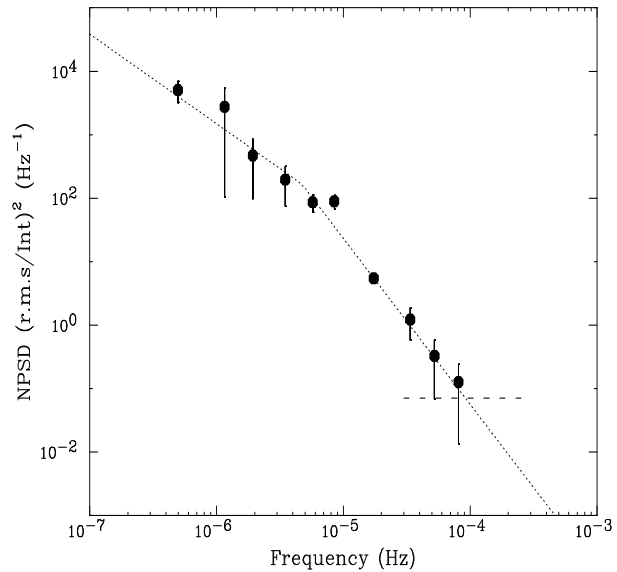,width=8.0cm,height=7.5cm}
\caption[h]{Normalized PSD (NPSD)
calculated from the light curves in Figure~1.
The dotted line shows the best-fit broken power-law function. Full details
 are given in the text.}
\label{fig3}
\end{figure}

\subsection{Structure Function}

The structure function (hereafter SF) is a numerical technique
similar to the traditional PSD, but which has some
advantages for dealing with highly
undersampled data. Since the SF is less affected by data gaps
in light curves (e.g., Hughes, Aller, \& Aller 1992), it may be
a useful estimator for our study. The definitions of SFs and their
properties are given in Simonetti, Cordes, \& Heeschen (1985).
The first order SF is defined as
\bq0
{\rm SF}(\tau) = \frac{1}{N}\sum[a(t) - a(t+\tau)]^2,
\label{equation:1-1}
\eq
where $a(t)$ is a point in the time series (light curve) ${a}$, and the
summation is made over all pairs separated in time by $\tau$.
The term $N$ is the number of such pairs.

The SF is closely related to the PSD. If the structure function has a
power-law form, $SF(\tau)$ $\propto$ $\tau^{\beta}$ ($\beta \ge 0$),
then the power spectrum has a distribution $P(f) \propto f^{-\alpha}$,
where $f$ is the frequency and $\alpha \simeq \beta + 1$.
Obviously such an approximate relation is violated when $\beta$ is close to
zero, since both $\alpha$ and $\beta$ approach zero for
white noise. Nevertheless, the SF gives a crude but convenient estimate of
the corresponding PSD distribution especially for red-noise type PSDs
(e.g., Paltani et al. 1997; Cagnoni, Papadakis, \& Fruscione 2001).

In Figure~4, the SF is calculated from the light curve presented in
Figure~1.  We used all the data as  a ``continuous'' observation in
order to probe the variability on timescales as long  as possible.
$RXTE$ (PCA) light
curves binned in 5760~sec intervals are used for the calculation.
The resulting SFs are normalized by the square of the mean fluxes and
are binned at logarithmically equal intervals.
The measurement noise (Poisson errors associated with flux uncertainty)
is subtracted as twice the square of Poisson errors on the fluxes
(and the noise level is given as the dashed line in the figure).

As suggested by the NPSD in Figure~3, the resulting SFs are
characterized by a steep increase ($\beta \simeq 1.5$) in the time
region of $0.1 \le \tau \le 3$~days, above which the SFs show a
significant roll-over ($\beta \simeq 0.3$).
A gradual rise continues up to a time
lag of $\tau \sim 30$~days, where a significant ``bump'' appears.
This bump, and the ``wiggling'' features at the longest
time-scales may be artefacts caused by the sparse sampling of the
light curve (see Cagnoni, Papadakis, \& Fruscione 2001; Kataoka et
al.\ 2001 for a detailed discussion).
The number of pairs ($N$ in equation (2)) decreases as the time separation
$\tau$ increases and, accordingly, the uncertainty becomes larger.

The most rigorous study of the nature of this ``bump'' would be obtained by
simulating the light-curves characterized by a certain SF, and filtering
them with the same window as the actual observation. The resulting SF could
then be compared with that adopted for the simulations.
However, such an estimate is only possible
when we already know the $true$ ``bump'' structures of the system.
The study of such structures is of interest in the light of
claims for $quasi$ $periodic$ $oscillation$ (QPO) on time-scales
of $\sim$\,100~day for other blazars (e.g., Rieger \& Mannheim 2000;
Abraham 2000), but is beyond the scope of this paper.
Here we limit discussion of the SF to time-scales shorter than 10~days,
where $N$ (see equation (2))
is large and the PSD and SF show good agreement.

In Figure~4, we also show the SF of Mrk~421
calculated from $ASCA$ observations over 5~years for comparison 
(solid line; reproduced from Kataoka et al.\ 2001). Mrk~421 is an  
``orthodox'',  and best studied blazar in the X-ray band. 
Two important characteristics are seen in the figure;
(i) the structure functions of both Mrk~421 and 3C~273 show a roll-over
at $1 \le t_{\rm var} \le 10$~days, and
(ii) the normalized SF, which is proportional to the square of variability
amplitude, is 1--2 orders of magnitude $smaller$ for 3C~273  than for
that of Mrk~421.

\begin{figure}
\psfig{file=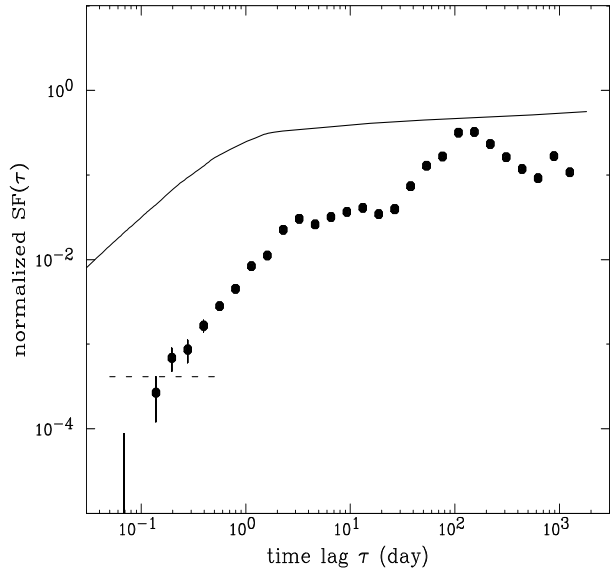,width=8.0cm,height=7.5cm}
\caption[h]{The structure function calculated from the light curves of
 Figure~1. The solid line shows the SF of the HBL Mrk~421 to compare with
 3C~273 (QHB). Full details are given in the text.}
\label{fig4}
\end{figure}

\section{Spectral Study}
\subsection{X-ray Spectral Evolution: power-law component}

Time variability of blazars is generally accompanied by significant
spectral changes, which provide direct information about the
acceleration, cooling and injection of electrons in the relativistic
jet. Notably, such spectral information gives independent, and/or
alternative information from the variability studies
presented in $\S$ 3.
In this section, we investigate the photon spectra evolution of 3C~273
for each observation conducted from 1996 to 2000.
We first divided the total
exposure into one-orbit (5760~sec) intervals to investigate the most
rapid evolution in photon spectra as possible.
Spectral fits have been performed individually for
261 segments over the four years of observations.
A single power-law function  and
a photoelectric absorbing column $N_{\rm H}$ fixed at the Galactic
value ($1.79 \times 10^{20}$\,cm$^{-2}$; Dickey \& Lockman 1990)
represent most of the spectra well in the 2.5$-$20~keV data. The
reduced $\chi^2$ for 251 segments (out of 261) ranges from 0.4 to 1.4 for 45
degrees of freedom, which corresponds to $P(\chi^2) \ge 5\%$.

\begin{figure*}
\psfig{file=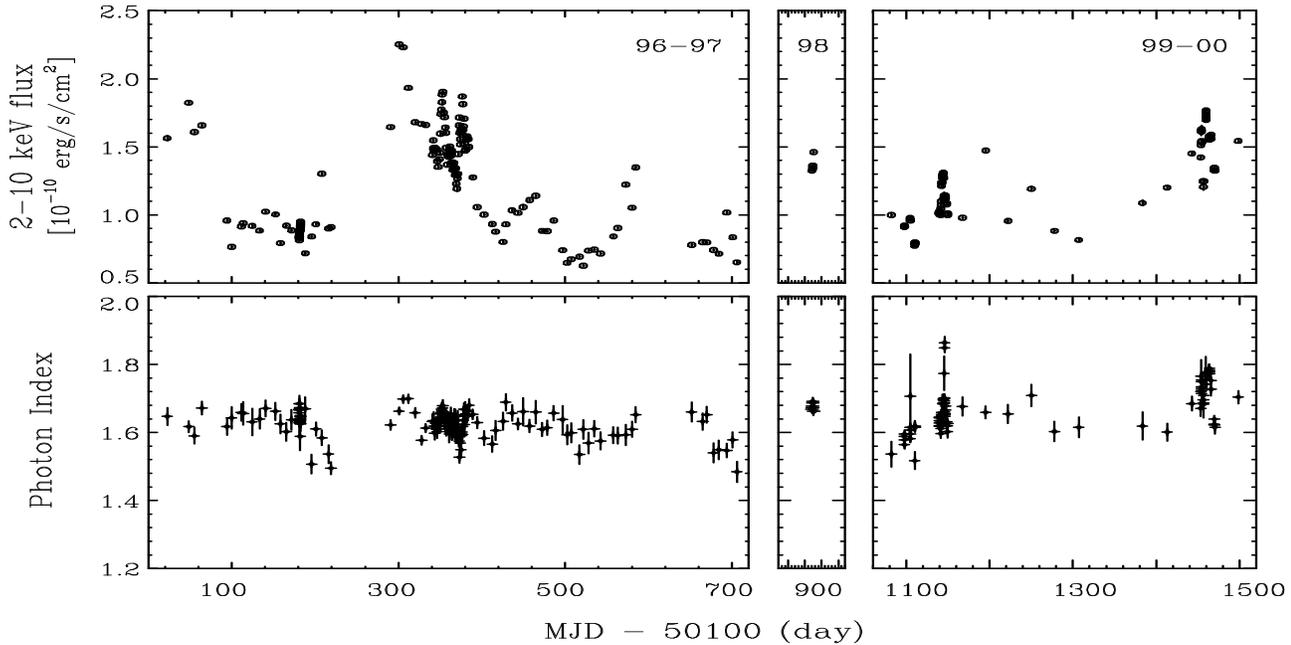,width=17cm,height=8.5cm}
\caption[h]{The spectral evolutions of 3C~273 as a function of time.
$upper$; changes in the 2$-$10~keV flux.  $lower$; changes in the differential
 photon index $\Gamma$.}
\label{fig5}
\end{figure*}

\begin{figure*}
\hspace{5mm}\psfig{file=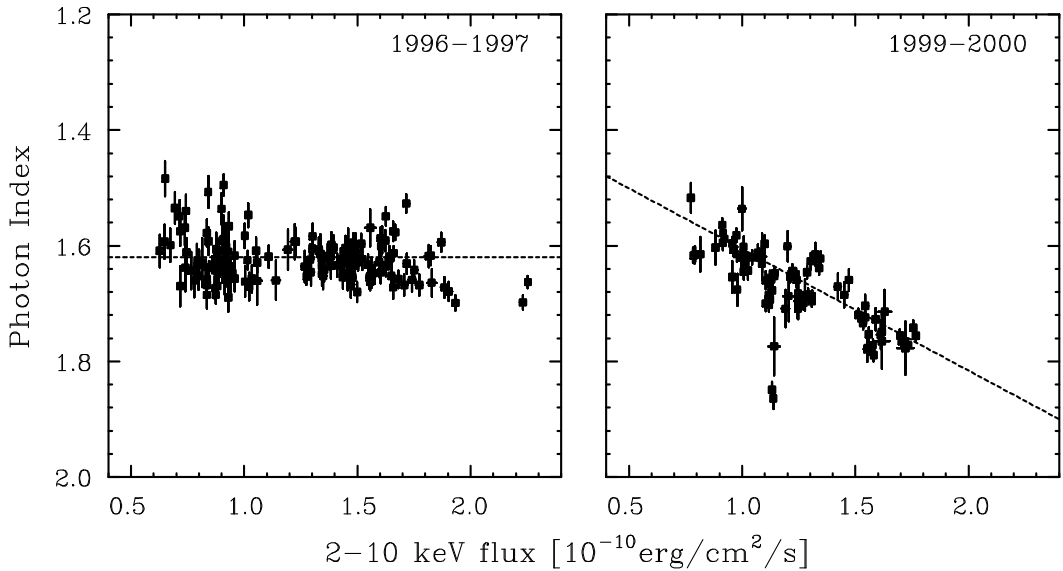,width=14cm,height=7.5cm}
\caption[h]{Correlations between the 2$-$10~keV flux and the photon
 index $\Gamma$. $left$; those for the 1996$-$1997 season (P10330, P10354, P20349);
$right$; those for the 1999$-$2000 season (P40176, P40177).}
\label{fig6}
\end{figure*}

We summarize the results of power-law fitting in Figure~5.
The upper panel shows the changes in the 2$-$10~keV flux and the evolution
of differential photon index ($\Gamma$) determined in the 2.5$-$20~keV
band.  We are aware that $RXTE$ is not sensitive to photons in the
2.0$-$2.5~keV. Nevertheless, the 2$-$10~keV flux is specifically selected
because most previous works have defined the X-ray flux in this energy
band (e.g., Turner et al. 1990; Yaqoob et al. 1994; Haardt et al. 1998).
The 2$-$10~keV flux changes dramatically during the observations in
1996$-$2000; from $2.25 \times 10^{-10}$~erg/cm$^{2}$/s at the brightest
(MJD~50400.8) to  $0.62 \times 10^{-10}$~erg/cm$^{2}$/s at the
faintest (MJD~50621.7).
During the 1996$-$1997 season, the hard X-ray spectral photon
indices stayed almost constant, with a
differential photon index of $\Gamma = 1.6 \pm 0.1$.
Larger spectral variation is seen in the 1999$-$2000 season,
where $\Gamma$ varies from 1.5 to 1.9.

The spectral evolution is more clearly seen  in Figure~6
where the correlation between the fluxes and the photon indices are
separately plotted for the 1996$-$1997 and 1999$-$2000 seasons.
For the 1996$-$1997 observations, the spectral shapes do not change
significantly  ($\Gamma$$\simeq$1.6) despite large flux variations.
The correlation factor between the flux and photon index is
$R_{\rm F \Gamma} = -0.19$.  A stronger correlation has been found
for 1999$-$2000 observations, where the spectral index becomes $softer$ when
the source is brighter ($R_{\rm F \Gamma} = -0.73$). 

\subsection{The Soft Excess and Broad Line Feature}

As mentioned above, 10 out of the 261 photon spectra cannot
be fitted by a simple power-law function, in the sense that
$P(\chi^2) \le 5\%$. There are two different types of departure
from a power-law spectra; (i) a significant soft excess below 4~keV, or
(ii) a broad-line feature seen at 5$-$6~keV (in the observer's frame).

Figure~7 shows examples of $\nu$$F_{\nu}$ photon spectra for each
case. The left panel shows a concave feature with an apparent break
at 4~keV (case (i)). A simple power-law fitting to the data gives
$\chi^2 = 1.4$ for 44~dof, where $P(\chi^2) = 4\%$. Spectral fits with
a broken power-law model significantly improve the goodness of the fit,
as summarized in Table~2.  Note that, below this break energy,
the photon spectrum shows a very steep power-law index of
$\Gamma_{\rm L} = 2.27 \pm 0.19$. A similar soft excess has been found in
the photon spectra of the segment closest in time (MJD~51246.15--51246.16),
however, the statistical significance is relatively low and the simple
power-law function gives a ``good'' fit ($P(\chi^2) \ge 27\%$).

\begin{table*}
\centering
 \begin{minipage}{150mm}
  \caption{Spectral fits to 3C~273 with a broken power-law function.}
  \begin{tabular}{@{}lclcccl@{}}
\hline
   Obs.date   & $\Gamma_{\rm L}$\footnote{Photon index of the low-energy
   power-law component.} & $E_{\rm brk}$\footnote{The break energy.}  &
$\Gamma_{\rm H}$\footnote{Photon index of   the high-energy    power-law
   component.} &  2$-$10 keV Flux &  reduced $\chi^2$ (dof) &
   $\Delta$$\chi^2$\footnote{The decrease in $\chi^2$ when the break is added to the simple power-law function.}\\
   (MJD)  &  & (keV) &  & (10$^{-10}$ erg/cm$^{2}$/s) &  & \\
\hline
  51246.10$-$12 & 2.27$\pm$0.19 & 4.02$\pm$0.37 & 1.79$\pm$0.02 & 1.17 & 0.86(42) & 25.0\\
\hline
\end{tabular}
\end{minipage}
\end{table*}

\begin{table*}
\centering
 \begin{minipage}{150mm}
\caption{Spectral fits to 3C~273 with a power-law function plus
Gaussian line model.}
\begin{tabular}{@{}lllllcll@{}}
\hline
   Obs.date   & $\Gamma$\footnote{Photon index of a power-law
   component.} & $E_{\rm Fe}$\footnote{Line center energy. Intrinsic
 line width is fixed at 0.8 keV.}
   & $I_{\rm Fe}$\footnote{Intensity of a line.} &
   EW\footnote{Equivalent width of a line.} & 2$-$10 keV Flux &  $\chi^2$
   (dof) & $\Delta$$\chi^2$\footnote{The decrease in $\chi^2$ when the Gaussian emission line is added to a simple power-law function.}\\
   (MJD)   &  & (keV) & (10$^{-4}$/cm$^{2}$/s) &
   (eV) & (10$^{-10}$ erg/cm$^{2}$/s) &  & \\
\hline
  50281.14$-$17 & 1.63$\pm$0.01 & 6.32$\pm$0.19  & 2.40$\pm$0.88 & 187$\pm$68 & 0.82 & 0.75(43) & 30.8\\
  50282.01$-$05& 1.62$\pm$0.01 & 5.82$\pm$0.17  & 2.91$\pm$0.98 &  184$\pm$62 & 0.88 & 0.78(43) & 37.5\\
  50282.14$-$17& 1.65$\pm$0.02 & 6.10$\pm$0.30  & 1.65$\pm$1.02 &  108$\pm$67 & 0.91 & 1.18(42) & 11.7\\
  50282.28$-$31 & 1.62$\pm$0.01 & 6.24$\pm$0.20  & 2.44$\pm$0.90 & 167$\pm$62 & 0.92 & 0.85(43) & 30.3\\
  50282.35$-$38 & 1.65$\pm$0.02 & 6.43$\pm$0.19  & 2.87$\pm$1.02 & 206$\pm$73 & 0.92 & 1.21(43) & 32.1\\
  50300.90$-$91 & 1.62$\pm$0.02 & 7.11$\pm$0.27  & 2.53$\pm$1.23 &  210$\pm$102 & 0.93 & 1.02(30\footnote{Fitting was performed over 2.5--15 keV due to low photon statistics.}) & 16.5\\
  50400.82$-$83 & 1.65$\pm$0.01 & 6.05$\pm$0.23  & 4.98$\pm$2.17 &   132$\pm$58 & 2.24 & 1.04(43) & 21.4\\
  50454.33$-$34 & 1.63$\pm$0.02 & 6.08$\pm$0.26  & 5.44$\pm$2.65 &   189$\pm$92 & 1.74 & 1.09(43) & 16.8\\
  50988.32$-$36 & 1.65$\pm$0.01 & 6.10$\pm$0.21  & 2.54$\pm$1.04 &
   117$\pm$48 & 1.31 & 1.09(43) & 25.5\\
\hline
\end{tabular}
\end{minipage}
\end{table*}

The broad-line features, as illustrated in Figure~7 ($right$), are
observed 9 times during the 1996$-$1998 observations. The results of spectral
fits to a power-law plus Gaussian line model are given in Table~3.
The line center energy is distributed between 5.8 and 7.1~keV in the quasar's
rest frame. We cannot determine the exact shape and width of this broad
line due to the limited sensitivity of the PCA on-board $RXTE$.
Following Yaqoob $\&$ Serlemitsos (2000),
we thus fixed the intrinsic line width at
$\sigma_{\rm Fe} = 0.8$~keV. The equivalent width of the line ranges
from 108 to 210~eV although uncertainties are large.
Taking these uncertainties into account, it seems premature to
identify this structure as a broad Fe $K_{\alpha}$ line only using the
$RXTE$ data. Comparison with previous works,
especially taken by instruments on-board other satellites,
is necessary for confirmation.

The broad line feature was firstly reported by Turner et al.\ (1990)
using the data taken by $EXOSAT$ and $Ginga$.
$RXTE$ detections of the Fe $K_{\alpha}$ line from 3C~273 have been
reported by Yaqoob $\&$ Serlemitsos (2000) for the observations in
July 17$-$18, 1996 (MJD 50181$-$50182) and June 24$-$26, 1998
(MJD 50988$-$50990). Importantly, these observations were
conducted simultaneously with
$ASCA$. $ASCA$ carried two different types of spectrometers, the Solid-state
Imaging Spectrometer (SIS; Yamashita et al.\ 1997) and the Gas Imaging
Spectrometer (GIS; Ohashi et al.\ 1996), both of which have
much better line sensitivity than the $RXTE$ PCA below 10~keV. Their results
clearly indicate the existence of a broad-line feature around 6~keV of
$EW \sim 100$~eV (Figure~2 in Yaqoob $\&$ Serlemitsos 2000). Our analysis
thus confirms the existence of broad line features for two observations
reported in Yaqoob $\&$ Serlemitsos (2000), and also adds new evidence from
other observations in 1996$-$1997.

It would be of interest to search for
correlations between the presence of the line
and the 3C273 flux, however, the data available do not allow us to
perform a convincing analysis of the line variability or of its
relationship with the continuum variations.
A future study along these lines will be very important for understanding
where the cold matter
emitting the fluorescence line is located with respect to the primary
source.

\begin{figure}
\psfig{file=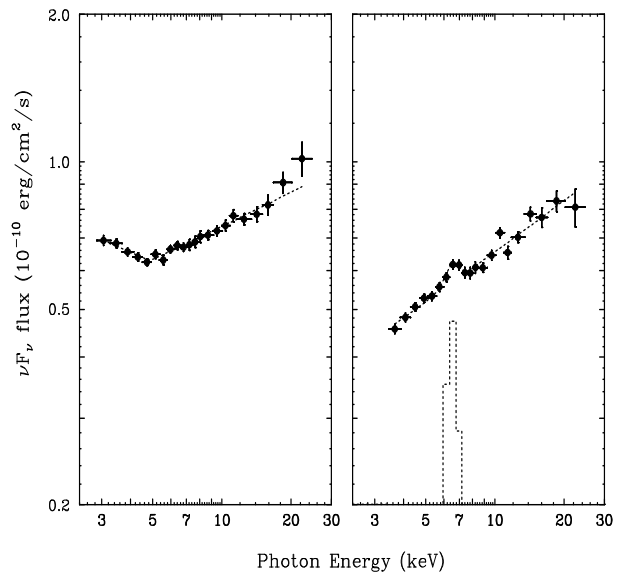,width=8.0cm,height=7.5cm}
\caption[h]{Examples of spectral fits to 3C~273 photon spectra when
(a) a soft excess and (b) a broad line feature are present.
The spectra are from data segments MJD 51246.10--51246.12
and MJD 50282.14--50282.17, respectively.
}
\label{fig7}
\end{figure}

\subsection{Multi-frequency Spectrum}

In order to understand the origin of hard X-ray emission from 3C~273 in
more detail, we constructed a $\nu L_{\nu}$ multi-frequency spectrum
adding our new $RXTE$ results. The resultant multi-frequency
spectrum in the quasar's rest frame ($z = 0.158$) is shown in Figure~8.
Crosses show $RXTE$ data when the source was in the brightest
state (MJD~50400.82$-$50400.83), whereas the filled circles show the data
for the faintest state (MJD~50621.67$-$50621.69). Open circles and
squares are archival data from literature; NED data base (radio to
UV), $ROSAT$ (soft X-ray; Staubert 1992), OSSE, COMPTEL and EGRET
on-board $CGRO$ ($\gamma$-ray; von Montigny et al.\ 1993; 1997).
Three different peaks are apparent in the spectra;
at low-energies (LE: radio to optical), high-energies (HE: X-ray to
$\gamma$-ray), with the $BBB$ (optical to soft X-ray)
in between.

To estimate the peak frequencies and luminosities of each component
quantitatively, we fitted the spectral energy distribution (SED)
with separate polynomial functions of the form;
log($\nu$$L_{\nu}$) = $a$ + $b$ log$\nu$ +
$c$ (log$\nu$)$^2$, where $a$, $b$, $c$ are constants.
Similar polynominal fits were applied in Comastri, Molendi \& Ghisellini
(1995) for various blazars assuming cubic functions. In the case
of 3C~273, however, there are only negligible differences between
the quadratic and cubic functions and hence,
we adopt the simpler quadratic form in the fitting.
The results of quadratic fits are summarized in Table~4, where
$L_{\rm p}$ is the $peak$ luminosity in the $\nu$$L_{\nu}$ space, and
$L_{\rm tot}$ is the $integrated$ luminosity over that
frequency range.

\begin{figure*}
\hspace{14mm}
\psfig{file=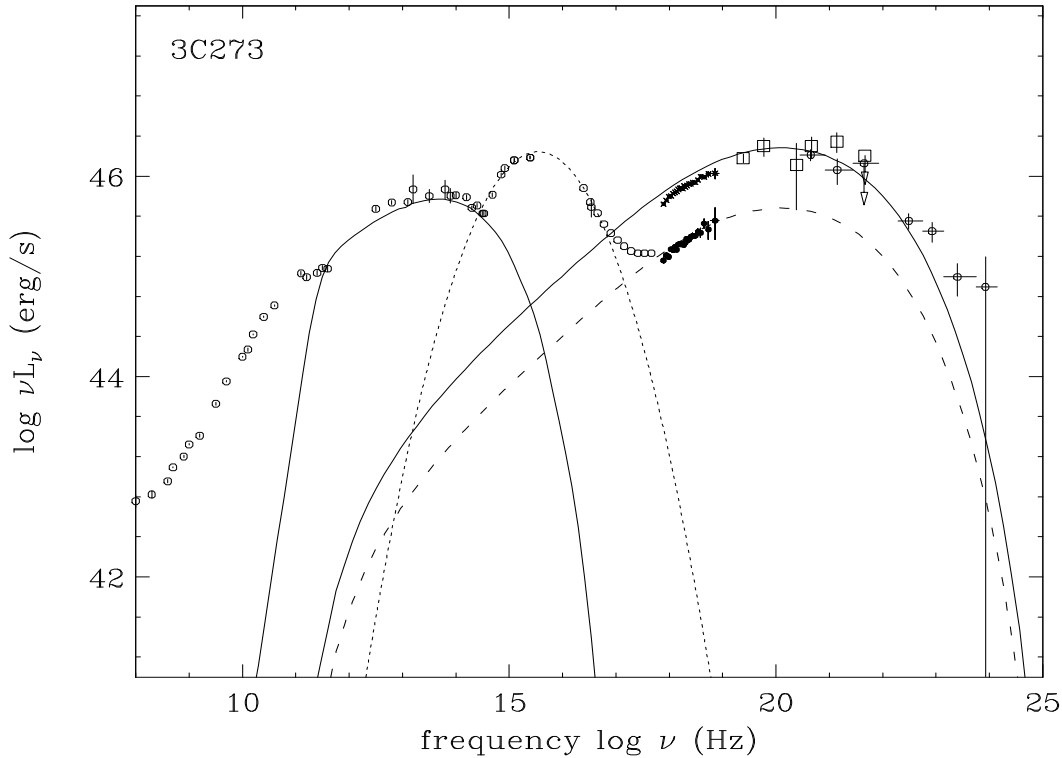,width=14cm,height=10cm}
\caption[h]{The multi-frequency spectra of 3C~273. $Open$ $circles$ and
$squares$; archival data from the literature. $Crosses$; $RXTE$ data when the
source was the brightest, MJD~50400.8 (this work).
$filled$ $circles$; $RXTE$ data when the  source was the faintest,
MJD~50621.7 (this work). The solid line shows the SSC model for parameters
$B = 0.4$~G, $\delta = 10$, $\gamma_{\rm max}$ = 2000, and
$R$ = 2.5 $\times$ $10^{16}$~cm. The dashed line shows a model for
parameters  $B$ = 0.4~G, $\delta$ = 10, $\gamma_{\rm max}$ = 2000,
and $R$ = 5.0 $\times$ $10^{16}$~cm with moderate electron number
 density. All parameters are determined
self-consistently as discussed in the text. The dotted line shows a
quadratic fit to the $BBB$ to guide the eye.}
\label{fig8}
\end{figure*}

It must be noted that the most of the data in Figure~8 were not obtained
simultaneously.  It is clear from a number of observations conducted
at various times and at various wavelengths that 3C~273 is
variable from radio to $\gamma$-ray energies
(see von Montigny et al. 1997; Courvoisier 1998). Even for this
particularly well-sampled and bright object, it has not yet been possible
to obtain an actual ``snap-shots'' covering the  total energy band pass.
Fortunately, the 30-year monitoring of 3C~273 reveals that the variability
amplitude is factor of 4 at most energy bands
(T\"{u}rler et al. 1999).
This is relatively small when plotted in the
log($\nu$)--log($\nu$$L_{\nu}$) plane, and we thus
believe our discussion below is not affected significantly by the
non-simultaneity of the data. We will comment on the effect of changes
in physical parameters in $\S$5.3.

\begin{table}
\centering
 \begin{minipage}{85mm}
  \caption{Quadratic fit results of 3C~273 SED.}
  \begin{tabular}{@{}lccc@{}}
\hline
   components    & $\nu_{\rm p}$ (Hz) & $L_{\rm p}$ (erg/s) & $L_{\rm tot}$ (erg/s)\\
\hline
       LE        &  10$^{13.5}$    & 10$^{45.8}$ & 10$^{46.7}$  \\
       HE\footnote{X-ray data taken during the brightest state are used
   for the fitting (MJD~50400.8; crosses in Figure 8.)} &  10$^{20.0}$    & 10$^{46.2}$ & 10$^{47.1}$  \\
       BBB        &  10$^{15.5}$    & 10$^{46.2}$ & 10$^{46.8}$  \\
\hline\\
\end{tabular}
\end{minipage}
\end{table}

\section{Discussion}
\subsection{Time Variability}

In $\S$3, we have estimated the X-ray variability time-scale 
of 3C~273 using two different methods. We found that a roll-over appears 
at a time-scale of $t_{\rm var}$$\sim$\,3~day, which is similar to 
the X-ray variability time-scale of Mrk 421 
($t_{\rm var}$$\sim$\,1~day; Figure 4). 
However, a clear difference was found in the $amplitude$ of the 
variability (normalization in Figure 4). In fact, the X-ray 
flux of Mrk 421 sometimes changes by more than a factor of 2 in a day (e.g.,
Takahashi et al. 2000), whereas 
flux variations of only 20$-$30$\%$ are detected for 3C~273 (Figure 2). 
Two possibilities may be considered to account for such difference 
in variabilities.

We first assume 3C~273 as a ``pure'' blazar, i.e., most of
X-rays originate in beamed jet emission. 
Such an assumption may be valid since most of the time 
the X-ray photon spectra are well represented by a power-law function 
similar to other blazar-type objects. 
As many authors have suggested, variability in blazars may 
occur in a variety of ways, e.g., changes in acceleration rate of 
electrons, magnetic field strength and/or beaming factor. 
Most convincing scenario which may reproduce the observation well is
that electrons are accelerated to higher energy during the flare than
usual (e.g., Mastichiadis \& Kirk 1997; Kataoka et  al. 2000). If this 
is the case, the different variability amplitudes may result from
different physical processes for the X-ray production.

As expected from the multi-frequency spectra, the X-ray photons are
thought to be emitted from the $highest$ energy electrons via the 
synchrotron process for Mrk 421 (Takahashi et al. 2000), 
whereas the inverse Compton emission from the $low$-$energy$ electrons
are responsible for the case of 3C~273 (see Figure 8). Since the higher 
energy electrons cool faster ($t_{\rm cool}$ $\propto$ $\gamma^{-1}$, 
where $\gamma$ is the Lorentz factor of electrons; Rybicki \& Lightman
1979), it is a natural consequence that both the temporal and spectral 
evolutions are most pronounced in the highest end of electron
population, as observed in Mrk 421. This scenario may well 
explain the apparent difference of variability amplitudes in Mrk 421 
and 3C~273.

Alternatively, we next assume 3C~273 as ``half-Seyfert/half-blazar''. 
It is clear that at some epochs at least, the X-ray 
flux has components that are $not$ related to the beamed jet 
emission and that are observed to have fluxes similar to those
discussed. This is particularly true of 
the soft excess and the broad line feature discussed in $\S$4.2. 
Thus we may have observed a combination of non-thermal X-ray
emission from the beamed jet, as well as the isotropic
thermal emission from the accretion disk at the same time. 
The two types of emissions cannot be discriminated only from the X-ray data,
however, variability in the lower-energy band, in particular the $BBB$, 
provides an important hint.

The variability of the soft excess has been monitored for a long time. 
For example, Paltani, Courvoisier, \& Walter (1998)  reported both
optical and ultraviolet observations of 3C~273 covering 
the lifetime of the $IUE$ satellite. They investigate the variability 
time-scales of $BBB$ using a structure function analysis. 
While resulting SFs are characterized by a steep increase as for the 
X-ray band ($\beta$$\sim$1.5; Figure 4), a plateu appeared at much
$longer$ time scale ($\tau$ $\sim$ 0.5 year; but see Courvoisier et al. 
1988 for an exceptionally rapid flare in the optical band).

If $\tau$ $\sim$ 0.5 year is a ``typical'' variability time-scale of the 
non-beamed isotropical component in 3C~273, this would produce a
``quasi-steady'' underlying component to the ``rapidly variable'' 
beamed jet emission. Such an offset may significantly reduce 
the amplitude of X-ray variability originated in the jet, but the observed 
time-scales (i.e., duration of a flare) would not be changed 
($\tau$$\sim$ 3 day). From this view-point, the difference of the variability 
amplitude in Mrk 421 and 3C~273 may be explained by the importance of 
underlying isotropic component over the beamed jet emission.

\subsection{Spectral Evolution}

In $\S$4, we showed that most 3C~273 photon spectra observed between
1996 and 2000 can be represented by a simple power-law function with
$\Gamma = 1.6 \pm 0.1$. This is consistent with previously published results.
Turner et al.\ (1990) suggest that the spectral index of 3C~273 in the
2$-$10~keV band is slightly harder than the ``canonical'' AGN index
of $\Gamma = 1.7$, though the difference seems to be very small (1.5 compared
to 1.7).  Such a difference, if confirmed, may be due to a different
physical origin of X-ray production.

In fact, X-ray photons are thought to be emitted from the Compton
scattering of disk photons for Seyfert-type AGNs, whereas accelerated
electrons in the jet produce non-thermal emission in the case of
blazars. In the standard shock scenario (Fermi acceleration), electrons
have a population $N(\gamma) \propto \gamma^{-2}$, where
$N(\gamma)$ is the number density of electrons per energy. Such
accelerated electrons emit photons with a spectrum of a 
form $\propto \nu^{-1.5}$,
through both the synchrotron and inverse-Compton processes
(e.g., Rybicki \& Lightman 1979). This is very close to  the
observational results.

If the hard X-ray photon index reflects the power-law index of
the accelerated electrons ($s$ of $\gamma^{-s}$),
only small spectral changes would be expected in the photon spectra
corresponding to the low-energy end of electron populations
(as for the X-ray photons of 3C~273). This is because low energy
electrons do not cool before leaving the emitting region, and hence keep
their ``original'' (injected) spectral information.
As suggested by Kirk, Rieger,
\& Mastichiadis (1998), changes in the electron injection rate could
produce flux variations, but the photon spectra do not evolve
significantly ($\propto \nu^{\frac{-(s+1)}{2}}$).
Such a scenario may well explain the spectral evolution observed in the
1996$-$1997 season,  where the photon index stays almost constant
($\Gamma$$\simeq$1.6) regardless of the source flux variations.

In 1999$-$2000, a different spectral evolution was found, with the 
spectra becoming steeper  (up to $\Gamma$$\sim$\,1.8) as the source becomes 
brighter. This may be related to the $underlying$ isotropic component 
emitted from the accretion disk or corona, as we have discussed in $\S$ 5.1.
It is interesting to note that the  exceptional ``concave'' photon 
spectrum was found only in this season (Figure 7 $left$). Such excess 
emission may be the hard-tail of the $BBB$, as suggested from a 
$ROSAT$ observation. Staubert (1992) reported a very steep X-ray
spectrum of $\Gamma_{\rm L}$ $\sim$ 2.6 in the $ROSAT$ bandpass 
(see also Figure~8), which is marginally consistent with our results 
($\Gamma_{\rm L} = 2.3 \pm 0.2$; Table~2).  
It is still unclear if the excess emission is really the hard tail of 
the $BBB$, as this feature was only clearly observed once, 
and $RXTE$ has relatively poor sensitivity in the soft X-ray band. 
However, it seems plausible that non-beamed, thermal emission and 
its reprocessed flux becomes more important in 1999-2000 season, 
compared to those in 1996-1998: future studies are necessary to confirm 
this.

The broad line feature is probably the Fe fluorescent line,
as already discussed in Turner et al.\ (1990; $EXOSAT$ and $Ginga$),
Cappi et al.\ (1998; $ASCA$) and Yaqoob $\&$ Serlemitsos (2000; $ASCA$
and $RXTE$). Interestingly, the Fe line is usually not visible in the
$RXTE$ observations; only 9 of 261 observations clearly
show line features, which is at odds with any thought that
the hard X-ray photons are emitted by the same
process with those in Seyfert galaxies. This suggests again that the X-ray
photons of 3C~273 are $ordinarily$ emitted from the relativistic jet, but
that $occasionally$ the emission from the accretion disk is visible.
(Although it should  be kept in mind  that the $RXTE$ PCA has a
relatively low sensitivity for the broad line feature, and that
observations with a more sensitive spectrometer may have detected the
line more frequently.) The visibility of the Fe fluorescent line may
depend on the balance between the jet emission, blue bump,
and/or Compton scattering of disk photons.
Future studies along these lines may be able to discriminate
between various models of disk/jet connection.

In summary, we conclude that the hard X-ray emission of 3C~273 may
occasionally contain a significant contribution from the accretion disk,
but most hard X-rays are likely to originate in inverse
Compton radiation from the relativistic jet.

\subsection{Multi-frequency spectrum}

Multi-frequency spectra provide information of physical quantities
relevant for jet emission, e.g., the magnetic field, the size of the
emission region, the maximum energy and the density of relativistic
electrons. Various authors have determined jet parameters in a self-consistent
manner for blazars (e.g., Mastichiadis $\&$ Kirk 1997; Kataoka et al. 2000;
Kino, Takahara, \& Kusunose 2002), but only a few attempts have been made
for 3C~273. We note again that part of the emission from 3C~273 
may $not$ be related to the beamed emission as discussed above. 
Nevertheless, we will consider this object as a ``blazar'' for the
following reasons; \\
(1) Contrary to many Seyfert galaxies (e.g., Mushotzky, Done \& Pounds
1993), there is no hint of a ``reflection hump'' corresponding to 
the Compton reflection of primary component.\\
(2) Although some exceptions have been found, the X-ray photon spectra
are generally well represented by a power-law function which smoothly 
connect to the $\gamma$-ray bands, as for other blazar-type objects. 

Von Montigny et al.\ (1997) examined three models
to reproduce the multi-frequency spectrum of 3C~273:
(i) the synchrotron self-Compton model (SSC model; e.g.,  Inoue \&
Takahara 1996), (ii) the external radiation Compton model
(ERC model; e.g., Sikora, Begelman, \& Rees 1994), and (iii) the proton induced
cascade model (PIC model; e.g., Mannheim \& Biermann 1992). Although
3C~273 is the one of the best sampled and best studied objects across
the entire electromagnetic spectrum, they conclude that the data were
still insufficient to discriminate between these models.

In the following, we thus adopt the simple one-zone SSC model to
describe the spectrum. The resultant physical
quantities provide a lot of feed-back to test the validity of our
assumptions.  We will also comment on the applicability of other
emission models, especially the ERC model,  in the appendix.
We do not consider the PIC model, in which the jet is composed of
$e^{-}$-$p$ plasma (baryonic) rather than $e^{-}$-$e^{+}$ plasma
(leptonic). Such a situation might be possible, but observationally, there
are no reasons for considering additional, baryonic emission to account
for the overall spectra (see Sikora \& Madejski 2001 for a detailed
discussion).

The jet parameters are tightly connected with the
observed quantities, particularly the $\nu_{\rm p}$ and $L_{\rm tot}$ of each
emission component (see $\S$4.3). In the simple model we adopt
here, the radiation is due to a homogeneous jet component
moving with a bulk Lorentz factor $\Gamma_{\rm BLK}$ =
(1 $-$ $\beta^2$)$^{-1/2}$ at an angle to the line of sight of
$\theta \simeq 1/\Gamma_{\rm BLK}$.

We first assume that the peak emission
($L_{\rm p}$) of the low-energy (LE) synchrotron component and the
high-energy (HE) inverse-Compton (SSC) components arise from the $same$
electron population with a Lorentz factor $\gamma_{\rm p}$.
The peak frequencies of LE and HE are related by
\begin{equation}
\nu_{\rm HE, p} = \frac{4}{3} \gamma_{\rm p}^2 \nu_{\rm LE, p},
\end{equation}
where $\nu_{\rm LE, p}$ = 10$^{13.5}$ Hz and $\nu_{\rm HE, p}$
= 10$^{20.0}$ Hz, respectively (Table~4). We thus obtain $\gamma_{\rm p}$ =
2.0$\times$10$^3$.
The synchrotron peak frequency, $\nu_{\rm LE, p}$, is given by
\begin{equation}
\nu_{\rm LE, p} \simeq 3.7\times 10^6 B \gamma_{\rm p}^2 \frac{\delta}{1 + z}  \hspace{3mm}{\rm Hz},
\end{equation}
where $B$ is the magnetic field strength and
$\delta$ = $\Gamma_{\rm BLK}^{-1}$($1-\beta$ cos $\theta$)$^{-1}$
$\simeq$ $\Gamma_{\rm BLK}$ is the relativistic beaming (Doppler) factor.
The magnetic field $B$ and its field densities are thus derived
\begin{equation}
B = 0.4\times(\frac{10}{\delta})\hspace{2mm}{\rm G},\hspace{2mm} U_B = 6.7\times 10^{-3}(\frac{10}{\delta})^{2}\hspace{2mm}{\rm erg}/{\rm cm}^3.
\end{equation}
Note that VLBI observations of 3C273 set a lower limit for the Lorentz
factor of $\Gamma_{\rm BLK}$\,$\ge$\,10, implying that
$\delta$\,$\sim$\, $\Gamma_{\rm BLK}$ $\sim$\,10 
if the angle to the line of sight of $\theta$\,$\simeq$\,
$1/\Gamma_{\rm BLK}$ (e.g., Pearson et al. 1981; Krichbaum et al. 2002).

The ratio of the synchrotron luminosity, $L_{\rm LE, tot}$,
to the inverse Compton (SSC) luminosity, $L_{\rm HE, tot}$  is
\begin{equation}
\frac{L_{\rm HE, tot}}{L_{\rm LE, tot}} = \frac{U_{\rm sync}}{U_B},
\end{equation}
where $U_{\rm sync}$ is the synchrotron photon energy density in the
co-moving frame of the jet. Using the observed value of
$L_{\rm HE, tot} = 10^{47.1}$~erg/s and
$L_{\rm LE, tot} = 10^{46.7}$~erg/s (Table~4), we obtain
\begin{equation}
U_{\rm sync} = 2.0\times 10^{-2}(\frac{10}{\delta})^{2}\hspace{3mm}
{\rm erg}/{\rm cm}^3.
\end{equation}

Assuming a spherical geometry for the emission region, the synchrotron
luminosity is given as
\begin{equation}
L_{\rm LE, tot} = 4\pi R^2 c \delta^4 U_{\rm sync}\hspace{3mm} {\rm erg/s},
\end{equation}
where $R$ is the radius of the emission region. This radius is estimated as
\begin{equation}
R = 2.5\times 10^{16} (\frac{10}{\delta})\hspace{3mm} {\rm cm}.
\end{equation}

Remarkably, the size of the region derived here corresponds to the variability
time-scale of $t_{\rm var} \simeq R/(c\delta) \simeq$ 1~day,
qualitatively consistent with those expected from the
temporal studies in $\S$3. Also,
if the jet is collimated within a cone of constant opening angle
$\theta \simeq 1/\Gamma_{\rm BLK}$, and the line-of-sight extent of the
shock is comparable with the angular extent of the jet, one expects that
the X-ray emission site is located at distance $D \sim 10^{17-18}$
~cm from the base of the jet.
This implies that the radiation discussed here is emitted
from the inner-most part of the jet, i.e., ``sub-pc-scale'' jet.
(This must be clearly distinguished from the ``large scale'' jet
described later in this section.)

The SSC spectrum is $self$-$consistently$ calculated assuming
the parameters derived above.
We assume an electron population of the form
$N(\gamma) \propto \gamma^{-s} {\rm exp}(-\gamma/\gamma_{\rm p})$,
where $s$ is set to be 2.
The solid line in Figure~8 shows the over-all model for the parameters
$R = 2.5 \times 10^{16}$~cm, $B = 0.4$~G, $\delta = 10$,
$\gamma_{p} = 2.0\times10^3$, while the dashed line shows the case when
the region size is larger by a factor of 2 with moderate electron number
density (and other parameters unchanged).
The whole spectrum is adequately represented, except
for the discrepancy in the radio band. Such discrepancies are common
for one-zone model fitting  of blazar SEDs, as
one-zone models cannot account for the low-energy emission,
which is thought to be produced in
a much larger region of the source (e.g., Marscher 1980).

The electron energy density, $U_e$, is calculated to be
$6.3 \times 10^{-1}$~erg/cm$^{3}$, where we set the minimum energy of
electrons to be $\gamma_{\rm min} = 1$.
The kinetic power of the electrons which emit the
observed photons is thus estimated to be
\begin{equation}
L_{\rm kin} \simeq \pi R^2 c \Gamma_{\rm BLK}^2 U_e \simeq
3.7\times10^{45}(\frac{\delta}{10})^2
\hspace{3mm}{\rm erg/s}.
\end{equation}

It is interesting to compare these radiative/kinetic luminosities to
discuss the ``power balance'' between the accretion disk and the
jet (see also Celotti, Padovani, \& Ghisellini 1997; Celotti, Ghisellini,
\& Chiaberge 2001 for the theoretical approach).
Assuming that the $BBB$ is emitted isotropically
from the accretion disk, the electro-magnetic luminosity in the accretion
disk is $L_{\rm disk} \simeq L_{\rm BBB, tot} = 10^{46.8}$~erg/s
(Table~4). The ``radiative power'' and ``Poynting power'' of the jet
in the observer's frame are  given as
\begin{eqnarray}
L_{\rm rad} \simeq \pi R^2 c \Gamma_{\rm BLK}^2 U_{\rm ph}
\simeq 5.0 \times 10^{44}(\frac{10}{\delta})^2 \hspace{3mm}{\rm erg/s} \\
L_{B} \simeq \pi R^2 c \Gamma_{\rm BLK}^2 U_{\rm B}
\simeq 3.9 \times 10^{43}(\frac{10}{\delta})^2 \hspace{3mm}{\rm erg/s},
\end{eqnarray}
respectively, where $U_{\rm ph}$ is the photon energy density in the jet
co-moving frame; $U_{\rm ph} \simeq U_{\rm sync} + U_{\rm SSC}$.
By comparing $L_{\rm disk}$, $L_{\rm rad}$, $L_B$, and
$L_{\rm kin}$, we derive the following properties of the ``power balance''
in sub-pc-scale jet; \\
(i) the kinetic power of $relativistic$ electrons amounts to only
    5$-$10\,$\%$ of the $BBB$ emission, which is thought to be
    radiated  in the vicinity of central black-hole.\\
(ii) the magnitudes of powers are
    $L_{\rm kin} \ge L_{\rm rad} \ge L_{B}$.

We finally comment on the large scale X-ray jet observed by
$Einstein$ (Harris \& Stern 1987), $ROSAT$ (R\"{o}ser et al. 2000) and
$Chandra$  (Sambruna et al.\ 2001; Marshall et al.\ 2001).
The X-ray jet is $\sim 8''$ (1$''= 2.4$~kpc at the
distance of 3C~273) long and has a knotty  morphology, starting
with a bright, resolved knot $\sim 13''$ from the core.
This ``kpc-scale jet'' is much fainter than the ``sub-pc-scale jet''.
Indeed, the integrated X-ray flux of the
kpc-scale jet is 6.9$\pm$0.6~nJy at 1~keV
(Marshall et al. 2001), which corresponds to $\sim$0.1$\%$ of the
X-ray flux emitted from the ``sub-pc-scale jet''.
The spectral energy distributions from radio to X-rays of the individual
knots show a variety of shapes. Sambruna et al. (2001) fit the
multi-frequency data of each knot with the ERC/CMB model, where the
cosmic microwave background (CMB) photons are up-scattered into the X-ray
band. For example, they obtain the best fit parameters for
region~A (see Sambruna et al.\ 2001 for definition) of electron index
$s = 2.6$, normalization $K = 8.1 \times 10 ^{-3}$~cm$^{-3}$,
magnetic field $B = 1.9 \times 10^{-6}$~G, region size $R =
5 \times 10^{21}$~cm, and Doppler factor $\delta = 5.2$.

Assuming these quantities, we estimate the kinetic power in the 10~kpc
scale jet as $L_{\rm kin, 10kpc}$ = 10$^{47.1}$($\frac{20}{\gamma_{\rm
min}}$)$^{0.6}$~erg/s, which is about 2 orders of magnitude larger
than that in the sub-pc-scale jet discussed above
($L_{\rm kin,sub-pc}$ $\sim$ 4$\times$10$^{45}$~erg/s; see equation (10)).
It should be noted that the kinetic power of electrons strongly depends
on both the minimum energy ($\gamma_{\rm min}$) and the power-law index
($s$) of the assumed electron population.
Part of the difference may be explained by different parameters
selected above, however, it seems difficult to completely explain
the discrepancy by this fact alone.
In fact, if we assume $\gamma_{\rm min} = 1$ for the 10~kpc
scale jet, as was assumed for the sub-pc-scale jet, $L_{\rm kin,10kpc}$
becomes $larger$, increasing the discrepancy. Similarly, an electron
population with $s = 2.0$ would minimize  $L_{\rm kin,10kpc}$, but such
parameters would not fit the observed SED so well. Future observations
of the 10~kpc scale jet with wider energy bands are eagerly awaited.

If the difference between the sub-pc and the 10~kpc scale jet powers is
confirmed, this may imply that $\sim$100 times the ``visible'' kinetic
energy is $hidden$ at the bottom of the jet. It should be noted
that the kinetic power discussed above takes account of only $relativistic$
electrons, in other words, the contribution from thermal electrons
as well as protons (either relativistic or cold) have completely been
neglected. These non-relativistic electrons and/or protons can not
contribute to the emission in sub-pc-scale jet.

Only a small fraction of them, probably less than 1\,$\%$ in number,
may be ``picked up'' by the shock acceleration process, to add to the
relativistic electron population. Such a low acceleration efficiency
may well be understood by the so-called $internal$ shock model as
discussed in Spada et al.\ (2001) and Kataoka et al.\ (2001).
Further discussion along this line is now in progress. For example,
Tanihata (2002) suggests that, in order to explain the observed
variability properties, the velocity difference  of the two colliding
shells must be rather small. For this case, the efficiency is estimated
to be as small as $\le$ 0.01 $\%$.

The enormous kinetic energy could be released for the first time at
large distances (e.g., $\ge$ 10 kpc) by completely different mechanisms
of energy dissipation. One such possibility is the
$external$ shock model, wherein shocks arise when outflowing jet
plasma decelerates upon interaction with dense gas clouds originating
outside the jet (e.g., Dermer \& Chiang 1998). The precise nature of the
required gas clouds is uncertain, but shock acceleration takes place more
efficiently than those for the $internal$ shock model.
Meanwhile, the different jet powers may imply that part of the kinetic
power carried by $protons$ in the sub-pc-scale
jet is efficiently transferred to
the $electrons$ in 10~kpc scale jet, though the mechanisms of energy
dissipation are completely uncertain. Future deep observations of large
scale jets by $Chandra$ and $Newton$ may be able to confirm our
suggestions about the power consumption in the 3C~273 jet.

\section{Conclusion}

We have analyzed the archival $RXTE$ data available for 3C~273
between 1996 and 2000. A total of 230 observations amounts to the net
exposure of 845~ksec over the 4 years.  This is the longest, and most
densely sampled, exposure for this object in the hard X-ray
band. Both the PSD and the SF show a roll-over with a time-scale on the
order of $\sim$\,3~days, although the lower-frequency (i.e., longer
time-scale) variability is still unclear. We found that the variability
time-scale of 3C~273 is similar to those observed in TeV gamma-ray 
emitting blazars, whereas the variability amplitude is an
order of magnitude smaller. Considering that the hard X-ray spectra of
3C~273 generally maintains a constant power-law shape with
$\Gamma \simeq 1.6 \pm 0.1$, beamed, inverse Compton emission inside
the jet is the most likely the origin of the X-ray/$\gamma$-ray emission.
Two kinds of exceptions have been found, which may be interpreted either
as the hard-tail of the $BBB$ or the emission from the accretion
disk $occasionally$ superposed on the jet emission. From
a multi-frequency analysis, we constrain the physical quantities
relevant for the jet emission. We argue that, (i) the kinetic power carried by
$relativistic$ electrons corresponds to only 5$-$10\,$\%$ of the disk
luminosity, and (ii) the various powers in the sub-pc-scale
jet are ranked $L_{\rm kin} \ge L_{\rm rad} \ge L_{B}$.
The connection between the sub-pc-scale jet and the 10~kpc scale jet
remains uncertain, but our work suggests that the most of the jet power
might be $hidden$ at the base of the sub-pc-scale jet, and effectively released
at the 10~kpc scale via a completely different mechanism of energy dissipation.

\section*{Appendix: Applicability and Validity of the SSC/ERC models}

Finally, we comment on an alternative scenario which may account for the
overall spectra of 3C~273, namely the external radiation Compton (ERC)
model. 

A number of quasars show complicated multi-frequency
spectra which cannot be readily fitted with a simple SSC model.
This is mostly because the $\gamma$-ray flux strongly dominates the
radiative output, and the $\gamma$-ray spectra are well above  the
extrapolation of the X-ray spectra (e.g., Ghisellini et al. 1998).
One explanation is that the SSC process dominates in the X-ray range, while
the ERC process dominates in $\gamma$-rays (e.g., Inoue \& Takahara
1996). Such a discontinuity/complexity cannot be seen for 3C~273
in Figure~8, suggesting that both the X-rays and $\gamma$-rays
are produced by the $same$ radiation process, i.e., the SSC process 
-- that is the main reason of why we apply the SSC model to 
the data in $\S$ 5.3.

However, this explanation is unable to reject other emission
models, in particular the ERC model. This is because most of data are 
obtained non-simultaneously, and we do not have an actual 
``snap-shot'' of the overall emission. In fact, the ERC model  provides 
similarly good fit for the case of 3C~273, as discussed in detail 
in von Montigny et al. (1997). In this appendix, we consider in what 
situation the SSC dominates the ERC process, and  what 
the expected impact would be if the ERC model is responsible for the 
overall emission from 3C~273.

Within the framework of ERC models, a number of possibilities exist 
for the origin of the seed photons which are up-scattered to
$\gamma$-rays. Dermer \& Schlickeiser (1993) suggest that the direct 
emission from the central core irradiates the emission region inside 
the jet, which may provide a sufficient number of seed photons. 
In the observer's frame, the isotropic luminosity of the central 
core is approximated by
\begin{equation}
L_{\rm core} \simeq L_{\rm disk} \simeq  L_{\rm BBB, tot} = 6.3\times10^{46} \hspace{3mm}  {\rm erg/s}
\end{equation}
(see Table~4).
In the co-moving frame of a jet with bulk Lorentz factor
$\Gamma_{\rm BLK}$, the photon energy density which is produced by
the core emission is given as
\begin{equation}
U_{\rm ext,in} = \frac{L_{\rm core}}{4 \pi d^2}\frac{\Gamma_{\rm BLK}^2(1 + \beta^2 - 2\beta)}{c} \simeq \frac{L_{\rm core}}{4 \pi d^2}\frac{1}{4 \Gamma_{\rm BLK}^2 c},
\end{equation}
where $d$ is the distance from the central core. By normalizing by typical
values of $\Gamma_{\rm BLK}$ and $d$, we obtain
\begin{equation}
U_{\rm ext,in} = 4.6\times10^{-3} (\frac{0.1\hspace{1mm}{\rm pc}}{d})^2 (\frac{10}{\Gamma_{\rm BLK}})^2 \hspace{3mm} {\rm erg/cm^3}.
\end{equation}
Comparing this with the synchrotron photon energy density,
$U_{\rm sync}$ = 2.0$\times$$10^{-2}$(10/$\delta$)$^2$
~${\rm erg}/{\rm cm}^3$ (see equation (7)), direct emission from the
central core cannot be a dominant source of seed photons for the ERC
process.  This is because the jet plasma is moving away from the central
core at highly relativistic speed ($\Gamma_{\rm BLK}$$\simeq$10), and hence
is strongly $red$-shifted as measured in the source co-moving frame.

The second candidate source of the external seed photons is the
``diffuse'' radiation field, i.e., a radiation field having a significant
non-radial component at large distance from the central engine.
Such a radiation component may be produced by scattering or reprocessing
of a portion of the central radiation by irradiated clouds and/or
the inter-cloud medium (Sikora, Begelman, \& Rees 1994). Importantly, all
photons directed inward toward the central source are $blue$-shifted,
and enhanced by a factor of $\Gamma_{\rm BLK}$ in the frame co-moving with the
jet. Assuming that a certain fraction $\tau$ ($<$ 1) of the central
luminosity is reprocessed into such a diffuse radiation field, we obtain
\begin{equation}
U_{\rm ext,out} = \frac{\tau L_{\rm core}}{4 \pi d^2}\frac{\Gamma_{\rm BLK}^2}{c}.
\end{equation}

If the SSC process is more likely origin of the X-ray/$\gamma$-ray
emissions, as we have assumed in $\S$ 5.3, inverse Comptonization of external
photons does $not$ dominate the synchrotron photons and
$U_{\rm sync} \ge U_{\rm ext,out}$ must hold.
This limits the fraction $\tau$ very tightly to
\begin{equation}
\tau \le 1.1\times10^{-4} (\frac{d}{0.1 \hspace{1mm}{\rm pc}})^2 (\frac{10}{\Gamma_{\rm BLK}})^4, 
\end{equation}
meaning that less than 0.01\,$\%$ of the $BBB$ luminosity 
could be reprocessed into the ``isotropic'' external radiation field 
in the case of 3C$~$273. One difficulty in understanding the overall photon 
spectrum of 3C$~$273 with the SSC model is why such a low
efficiency is achieved in this object -- unfortunately, we still do not 
have an answer to this question.

One may therefore consider the $opposite$ extreme, when the X-ray and
$\gamma$-ray emissions are dominated by the ERC process, rather than SSC 
emission. In such a situation, the ratio of the synchrotron luminosity,  
$L_{\rm LE, tot}$, to the inverse Compton (ERC) luminosity, 
$L_{\rm HE, tot}$, is  (see also equation (6))
\begin{equation}
\frac{L_{\rm HE, tot}}{L_{\rm LE, tot}} = \frac{U_{\rm ext,out}}{U_B}.
\end{equation}

The magnetic field density is estimated to be 
\begin{equation}
U_B \simeq \frac{L_{\rm LE, tot}}{L_{\rm HE, tot}} \frac{\tau L_{\rm core}}{4 \pi d^2}\frac{\Gamma_{\rm BLK}^2}{c}.
\end{equation}
The reprocessing factor $\tau$ is very uncertain, but it must be larger
than $10^{-4}$ if the ERC process dominates (see equation (17)).  
Normalizing $\tau$ with 10$^{-3}$,
and assuming  the observed value of $L_{\rm LE, tot}$,
$L_{\rm HE, tot}$, and $L_{\rm core} \simeq L_{\rm BBB, tot}$ (Table~3),
we obtain
\begin{equation}
B \simeq 1.4\times(\frac{\tau}{10^{-3}})^{\frac{1}{2}} (\frac{\Gamma_{\rm BLK}}{10}) (\frac{0.1\hspace{1mm}{\rm pc}}{d}) \hspace{3mm} {\rm G}.
\end{equation}
Thus in the framework of the ERC model, the overall emission of 3C~273 
may be reproduced if the magnetic field strength is a bit stronger than the SSC
case.

In both the SSC and ERC models, the low energy component is
thought to be produced by the same, synchrotron process. 
The synchrotron luminosity, $L_{\rm LE, tot}$ behaves as
\begin{equation}
L_{\rm LE, tot} \propto U_B U_e R^3.
\end{equation}
In order to explain the observed $L_{\rm LE, tot}$ with a $different$
magnetic field strength $B$ ($\simeq$0.4~G for the SSC and $\simeq$1.4~G
for the ERC model), the electron energy density $U_e$ must be varied
since $R$ is constrained to 10$^{16-17}$~cm from the observed
variability time-scale. Thus the $U_e$ of the ERC model could be smaller
by about an order of magnitude than we have assumed  for the  SSC model,
$U_e$ $\sim$ 5$\times$10$^{-2}$~erg/cm$^3$. This may lead to
an over-estimate of the kinetic power if the overall X-ray/$\gamma$-ray 
spectra are dominated by the ERC (see equation (10)).
However, what is important is that assuming either the SSC or 
the ERC model, our discussion and conclusions about the power balance in the
jet (see $\S$ 5.3) are not affected significantly.

\section*{Acknowledgments}

We greatly appreciate the referee, Dr J.~T. Courvoisier, 
for his helpful comments and suggestions to improve the manuscript.
This research has made use of the NASA/IPAC Extragalactic Database
(NED) which is operated by the Jet Propulsion Laboratory, California
Institute of Technology, under contract with the National Aeronautics
and Space Administration.


\begin{thebibliography}{99}
\bibitem{ab}Abraham, Z., 2000, A\&A, 355, 915
\bibitem{cag}Cagnoni, I., Papadakis, I. E., \& Fruscione, A., 
	2001, ApJ, 546, 886
\bibitem{capp}Cappi, M., et al. 1998, PASJ, 50, 213
\bibitem{cel1}Celotti, A., Padovani, P., \& Ghisellini, G., 1997,
	MNRAS, 286, 415
\bibitem{cel2}Celotti, A., Ghisellini, G., \& Chiaberge, M., 2001,
	MNRAS, 321, L1
\bibitem{coma}Comastri, A., Molendi, S., \& Ghisellini, G., 1995, MNRAS, 277, 297
\bibitem{cou98}Courvoisier, T. J.-L., 1998, Astron
	Astrophys Rev, 9, 1
\bibitem{cou88}Courvoisier, T. J.-L, et al. 1988, Nature, 335, 330
\bibitem{cou91}Courvoisier, T. J.-L., \& Clavel, J., 1991, A\&A, 248, 389
\bibitem{Cze}Czerny, B., 1994, in Multi-wavelength continuum emission of
	AGN, IAU Symp. 159, Eds Courvoisier and Blecha, Kluwer academic
        publishers
\bibitem{Der} Dermer, C. D., \& Schlickeiser, R., 1993, ApJ, 416, 458
\bibitem{Der2} Dermer, C. D., \& Chiang, J., 1998, NewA, 3, 157
\bibitem{Dic} Dickey, J. M., \& Lockman, F. J., 1990, ARAA, 28, 215
\bibitem{ghi} Ghisellini, G., et al. 1998, MNRAS, 301, 451
\bibitem{gran} Grandi, P., et al. 1997, A\&A, 325, L17
\bibitem{harr} Haardt, F., et a. 1998, A\&A, 340, 35
\bibitem{har} Harris, D, E., \& Stern, C, P., 1987, ApJ, 313, 136
\bibitem{Har} Hartman, R. C., et al. 1999, ApJS, 123, 79
\bibitem{haya} Hayashida, K., et al. 1998, ApJ, 500, 642
\bibitem{hug} Hughes, P. A., Aller, H. D.,\& Aller, M. F., 1992,ApJ,
	396, 469
\bibitem{inoue} Inoue, S., \& Takahara, F. 1996, ApJ, 463, 555
\bibitem{jah} Jahoda, K., et al. 1996, in ``EUV, X-ray, and
        Gamma-ray Instrumentation for Astronomy VII'', SPIE Proceedings,
        2808, 59
\bibitem{Kata3} Kataoka, J., et al. 2000, ApJ, 528, 243
\bibitem{Kata4} Kataoka, J., et al. 2001, ApJ, 560, 659
\bibitem{Kino} Kino, M., \& Takahara, F., \& Kusunose, M., 2002, ApJ,
	564, 97
\bibitem{Kirk} Kirk, J. G., Rieger, F. M. \& Mastichiadis, A. 1998, A\&A, 333, 452
\bibitem{Kri} Krichbaum, T. P., et al. 2002,  ASP Conf. Proc. 250 in press
eds. R.A. Laing and K.M. Blundell
\bibitem{Man} Mannheim, K., \& Biermann, P. L. 1992, A\&A, 253, L21
\bibitem{Mar}Marscher, A. P., 1980, ApJ, 235, 386
\bibitem{Marsh}Marshall, H. L., et al. 2001, ApJ, 549, L167
\bibitem{mastichiadis} Mastichiadis, A., \& Kirk, J. G., 1997, A\&A,
	320, 19
\bibitem{Mch}McHardy, I., et al. 1999, MNRAS, 310, 571
\bibitem{McN} McNaron-Brown, K., et al. 1995, ApJ, 451, 575
\bibitem{Muk}Mukherjee, R., et al. 1997, ApJ, 490, 116
\bibitem{Mus}Mushotzky, R, F., Done, C, \& Pounds, K, A., 1993, ARA\&A, 
	31, 717
\bibitem{Oha}Ohashi, T., et al. 1996, PASJ, 48, 157
\bibitem{Pal97}Paltani, S., Courvoisier, T. J.-L., Blecha, A., \&
	Bratschi, P., 1997, A\&A, 327, 539 
\bibitem{Pal98}Paltani, S., Courvoisier, T. J.-L., \& Walter, R., 
	1998, A\&A, 340, 47 
\bibitem{Pea}Pearson, T. J., et al, 1981, Nature, 290, 365
\bibitem{Rie}Rieger, F. M., \& Mannheim, K., 2000, A\&A, 359, 948
\bibitem{Rob}Robson, I., 1996, in ``Active Galactic Nuclei'' (New York: Wiley)
\bibitem{Ros}R\"{o}ser, H. -J., et al,  2000, A\&A, 360, 99
\bibitem{Ry} Rybicki, G. B., \& Lightman, A. P. 1979, in ``Radiative Processes
	in Astrophysics'' (New York: Wiley)
\bibitem{Sam} Sambruna, R. M., et al. 2001, ApJ, 549, L161
\bibitem{Shie} Shields, G. A., 1978, Nature, 272, 706
\bibitem{Sho} Sch\"onfelder, V., et al. 2000, A\&AS, 143, 145
\bibitem{Sik} Sikora, M., Begelman, M. C., \& Rees, M. J.,  1994, ApJ,
	421, 153
\bibitem{Sik2}Sikora, M., \& Madejski, G.,  in ``International Symposium
	on High Energy Gamma-Ray Astronomy'', Heidelberg, Eds
	F. Aharonian and H. V\"olk, 2001, AIP, 558, 275 (astro-ph/0101382)
\bibitem{simo} Simonetti, J. H.,  Cordes, J. M., \& Heeschen, D. S.,
	1985, ApJ, 296, 46
\bibitem{Spa} Spada, M., et al.  2001, MNRAS, 325, 1559
\bibitem{Stau} Staubert, R., 1992, MPE Report 235, 42
\bibitem{Swa} Swanenburg, B. N., et al. 1978, Nature 275, 298
\bibitem{Tak} Takahashi, T., et al. 2000, ApJ, 542, L105
\bibitem{Tan2} Tanihata, C., 2002, Ph D Thesis, University of Tokyo
\bibitem{Turl} T\"{u}rler, M., et al. 1999, A\&AS, 134, 89
\bibitem{Tur} Turner, M. J. L., et al. 1990, MNRAS, 244, 310
\bibitem{ulrich} Ulrich, M.-H., Maraschi, L., \& Urry, C. M. 1997, ARAA,
	35, 445
\bibitem{urry95} Urry, C. M., \& Padovani, P., 1995, PASP, 715, 803
\bibitem{ver} Vermeulen, R. C., \& Cohen, M. H. 1994, ApJ, 430, 467
\bibitem{von1} von Montigny, C., et al. 1993, A\&AS, 97, 101
\bibitem{von3} von Montigny, C., et al. 1997, ApJ, 483, 161
\bibitem{Yam} Yamashita, A., et al. 1997, IEEE Trans. Nucl. Sci., 44, 847
\bibitem{Yaq} Yaqoob, T., \& Serlemitsos, P., 2000, ApJ, 544, L95

\end{thebibliography}
\end{document}